\documentclass{opt2022}

\title[Quantization-based Optimization with Perspective of Quantum Mechanics]{Quantization-based Optimization with Perspective of Quantum Mechanics}

\optauthor{%
\Name{Jinwuk Seok} \Email{jnwseok@etri.re.kr}\\
\Name{Chang Sik Cho} \Email{cscho@etri.re.kr}\\
\addr Future Computing Research Division, Artificial Intelligence Research Lab. ETRI, Rep. Korea}

\usepackage{amsmath,amssymb,amsfonts}
\usepackage{algorithmic}
\usepackage{textcomp}
\usepackage{xcolor}
\def\BibTeX{{\rm B\kern-.05em{\sc i\kern-.025em b}\kern-.08em
    T\kern-.1667em\lower.7ex\hbox{E}\kern-.125emX}}

\usepackage{booktabs}
\usepackage{multirow}
\usepackage{multicol}
\usepackage{verbatim}
\begin{document}

\maketitle

\begin{abstract}
Statistical and stochastic analysis based on thermodynamics has been the main analysis framework for stochastic global optimization. 
Recently, with the appearance of quantum annealing or quantum tunneling algorithms for global optimization, we require a new research framework for global optimization algorithms. 
In this paper, we provide the analysis for quantization-based optimization based on the Schrödinger equation to reveal what property in quantum mechanics enables global optimization. 
We present that the tunneling effect derived by the Schrödinger equation in quantization-based optimization enables to escape of a local minimum.
Additionally, we confirm that this tunneling effect is the same property included in quantum mechanics-based global optimization.
Experiments with standard multi-modal benchmark functions represent that the proposed analysis is valid.
\end{abstract}

\section{Introduction}
Stochastic global optimization algorithms such as Simulated Annealing(SA) have represented outstanding performance in combinatorial optimization problems \cite{SA_83_01, GA_TSP_2017, Jiang_2007}.
However, when the complexity and size of a problem, such as a Traveling Salesman Problem(TSP) involving many cities (beyond 100 cities), are significantly huge, such a conventional algorithm shows a limitation of optimization performance \cite{Jinwuk_OTP_2022}.
Recently, the newest optimization algorithm, which applies quantization to the range space of an objective function, represented exceptional optimization performance in such an intricate problem \cite{Jinwuk_2022}.
Nevertheless, the dynamics of the quantization-based optimization are based on the analysis of the conventional stochastic global optimization, so it is difficult to realize the core component to reveal such superiority.
In this paper, we present the transformation from the Fokker-Plank equation, which describes the dynamics of the state transition probability in the quantization-based and stochastic global optimization, to the Schrödinger equation for the analysis based on quantum mechanics \citet{Hamacher_2006}. 
In addition, from experiments to compare the optimization performance concerning SA and Quantum Annealing(QA) (\citet{Kadowaki_1998, Santoro_2006}), we provide the validity of the quantization-based optimization algorithm in a general continuous objective function.
\section{Fundamental Definition and Assumption}  
First, we consider an objective function $f:\mathbf{R}^n \rightarrow \mathbf{R}^+$ with the unique global optimum $x^*$ such that $f(x^*) < f(x)$, for all $x, x^* \in \mathbf{R}^n$ and $x \neq x^*$. 
Further, we establish the following definitions and assumptions before beginning our discussion.
\begin{definition}
For $f \in \mathbf{R}$, we define the quantization of $f$ as follows:
\begin{equation}
f^Q \triangleq \frac{1}{Q_p} \lfloor Q_p \cdot (f + 0.5 \cdot Q_p^{-1} )\rfloor = \frac{1}{Q_p} (Q_p \cdot f + \varepsilon) = f + \varepsilon Q_p^{-1} 
\label{def_eq01}
\end{equation}
, where $\lfloor f \rfloor \in \mathbf{Z}$ is the floor function such that $\lfloor f \rfloor \in \max_{y} \{ y \in \mathbf{Z} | y \leq x, \; x \in \mathbf{R} \}$, $\varepsilon \in \mathbf{R}[-1/2, 1/2]$ is the quantization error, and $f^Q \in \mathbf{Q}$ is the quantization of $f$.
\label{def_01}
\end{definition}
In Definition $\ref{def_01}$, we establish the quantization parameter $Q_p \in \mathbf{Q}^+$ to be a monotone increasing function $Q_p : \mathbf{R}^{++} \mapsto \mathbf{Z}^{+}$ such that 
\begin{equation}
Q_p (t) = \eta \cdot b^{\bar{h}(t)} 
\label{def_eq02}    
\end{equation}
, where $\eta \in \mathbf{Q}^{++}$ denotes the fixed constant parameter of the quantization parameter, $b$ denotes the base, and $\bar{h} :\mathbf{R}^{++} \mapsto \mathbf{Z}^+$ denotes the power function such that $\bar{h}(t) \uparrow \infty \; \text{ as } \; t \rightarrow \infty$, for all $t \in \mathbf{R}^{++}$.
We assume that the quantization error defined in $\eqref{def_eq01}$ with a uniform distribution, according to the White Noise Hypothesis (WNH) \cite{Jimnez_2007}. 
This statistical assumption of the quantization error leads to the mean and the variance provided by the following proposition:
\begin{proposition}     
If the quantization error $\varepsilon_t \in \mathbf{R}^n$ satisfies the WNH, the mean and the variance of the quantization error at $t > 0$ is 
\begin{equation}
\forall \varepsilon^q_t \in \mathbf{R}, \quad
\mathbb{E}_{\mathbf{R}} Q_p(t) \varepsilon^q_t = 0, \quad
\mathbb{E}_{\mathbf{R}} Q_p^{-2}(t) {\varepsilon^q_t}^2 = Q_p^{-2}(t) \cdot \mathbb{E}_{\mathbf{R}} {\varepsilon^q_t}^2 = \frac{1}{12 \cdot Q_p^2 (t)} 
\label{prop_eq01}
\end{equation}
\label{prop_01}
\end{proposition}
Furthermore, we establish the notations of vector-valued derivatives as follows:
\begin{definition}
Suppose that $\{ \boldsymbol{e}_k \}_{k=1}^n$ denotes the set of basis vectors on an Euclidean vector space. 
We define the gradient, the divergence, and the Laplacian operation such that
\begin{equation}
\begin{aligned}
&\nabla_{\boldsymbol{x}} \triangleq \sum_k \frac{\partial}{\partial x_k} \boldsymbol{e}_k \in \mathbf{R}^n \quad
\nabla \cdot f(\boldsymbol{x}, \cdot ) \triangleq \sum_j \frac{\partial f}{\partial x_j} \in \mathbf{R}, \quad \because f:\mathbf{R}^n \rightarrow \mathbf{R}\\
&\Delta \triangleq \nabla \cdot \nabla = \sum_k \frac{\partial}{\partial x_k}(\sum_j \frac{\partial}{\partial x_j}) = \sum_k \sum_j \frac{\partial^2 }{\partial x_k \partial x_j} \in \mathbf{R}.
\end{aligned}
\label{def_eq03}        
\end{equation}
\end{definition}

Holding the above definition and assumption for the quantization error, we can establish the stochastic differential equation for the quantization-based optimization algorithm according to \cite{Jinwuk_OTP_2022}, as follows: 
\begin{proposition}     
For a given objective function $f: \mathbf{R}^n \rightarrow \mathbf{R}^+$, suppose that there exist the quantized objective functions $f^Q(\boldsymbol{x}_t), \; f^Q(\boldsymbol{x}_{t+1})$ evaluated from $\eqref{def_eq01}$, at a current state $\boldsymbol{x}_t \in \mathbf{R}^n$ and the following state $\boldsymbol{x}_{t+1} \in \mathbf{R}^n$ such that $f^Q(\boldsymbol{x}_t) \geq f^Q(\boldsymbol{x}_{t+1})$, for all $\boldsymbol{x}_{t+1} \neq \boldsymbol{x}_t$. We can obtain the stochastic differential equation of the state transition as follows: 
\begin{equation}
    d\boldsymbol{X}_t = - \nabla_{\boldsymbol{x}} f(\boldsymbol{X}_t) dt + \sqrt{C_q} \cdot Q_p^{-1}(t) d\boldsymbol{W}_t
\label{prop_eq02}
\end{equation}
, where $\boldsymbol{W}_t \in \mathbf{R}^n$ denotes a vector-valued standard Wiener process, which has a zero mean and variance with one, $\boldsymbol{X}_t \in \mathbf{R}^n$ denotes a random variable corresponding to $\boldsymbol{x}_t$.
\end{proposition}
Given the dynamics of the algorithm as $\eqref{prop_eq02}$, we can obtain the corresponding Fokker-Plank equation such that 
\begin{equation}
\partial_t \rho(\boldsymbol{x}, t) = \nabla \cdot (\nabla_{\boldsymbol{x}} f(\boldsymbol{x}) \rho(\boldsymbol{x}, t)) + \frac{1}{2} C_q Q_p^{-2}(t) \Delta \rho(\boldsymbol{x}, t)
\label{sec2_eq01}
\end{equation}
, where a state $(\boldsymbol{x}, t)$ instead of the random variable $\boldsymbol{X}_t$ at time t, and $\rho(\boldsymbol{x}, t) : \mathbf{R}^n \times \mathbf{R} \rightarrow \mathbf{R}[0,1]$ denotes a probability density function of the random variable $\boldsymbol{X}_t$. 

\section{Derivation of the Schrödinger Equation to Quantization-based Optimization}
\subsection{Derivation of the Schrödinger Equation from the Fokker-Plank equation for the Quantization-based Optimization}
For convenience, let a diffusion parameter $Q(t): \mathbf{R}^+ \rightarrow \mathbf{R}$ such that $Q(t) \triangleq C_q Q_p^{-2}(t)$.
Considering a log function to the probability density to $\boldsymbol{x}$ such as $\ln \rho(\boldsymbol{x}, t)$, we can calculate the gradient of the log function as follows:
\begin{equation}
\nabla_{\boldsymbol{x}}  \ln \rho(\boldsymbol{x}, t)
= \frac{\partial}{\partial \rho} \ln \rho(\boldsymbol{x}, t) \cdot \sum_{k} \frac{\partial \rho}{\partial x_k} \boldsymbol{e}_k
= \frac{1}{\rho(\boldsymbol{x}, t)} \nabla_{\boldsymbol{x}} \rho(\boldsymbol{x}, t).
\label{sec3_eq01}
\end{equation}
In addition, we establish a function $\mu(\boldsymbol{x}, t) : \mathbf{R}^n \times \mathbf{R} \rightarrow \mathbf{R}^n$ such that
\begin{equation}
\mu(\boldsymbol{x}, t) = \nabla_{\boldsymbol{x}}f(\boldsymbol{x}, t) + Q(t) \nabla_{\boldsymbol{x}} \ln \rho(\boldsymbol{x}, t)
\label{sec3_eq02}
\end{equation}
Substituting $\eqref{sec3_eq02}$ into $\eqref{sec3_eq01}$, we get 
\begin{equation}
\begin{aligned}
\mu(\boldsymbol{x}, t) = \nabla_{\boldsymbol{x}}f(\boldsymbol{x}) + Q(t) \frac{1}{\rho(\boldsymbol{x}, t)} \nabla_{\boldsymbol{x}} \rho(\boldsymbol{x}, t) 
\Rightarrow
\nabla_{\boldsymbol{x}}f(\boldsymbol{x}) \rho(\boldsymbol{x}, t) = \mu(\boldsymbol{x}, t) \rho(\boldsymbol{x}, t) - Q(t) \nabla_{\boldsymbol{x}} \rho(\boldsymbol{x}, t)
\end{aligned}
\label{sec3_eq03}
\end{equation}
Substituting $\eqref{sec3_eq03}$ into $\eqref{sec2_eq01}$, it leads 
\begin{equation}
\partial_t \rho(\boldsymbol{x}, t) = \nabla \cdot \mu(\boldsymbol{x}, t) \rho(\boldsymbol{x}, t) - \frac{1}{2} Q(t) \Delta \rho(\boldsymbol{x}, t).   
\label{sec3_eq04}
\end{equation}
Adding $\eqref{sec3_eq04}$ to $\eqref{sec2_eq01}$, we obtain the following equation:
\begin{equation}
\partial_t \rho(\boldsymbol{x}, t) = -\nabla \cdot v(\boldsymbol{x}, t) \rho(\boldsymbol{x}, t)    
\end{equation}
, where the function $v(\boldsymbol{x}, t) : \mathbf{R}^n \times \mathbf{R} \rightarrow \mathbf{R}^n$ is as follows: 
\begin{equation}
\begin{aligned}
v(\boldsymbol{x}, t)
= -\frac{1}{2}(\nabla_{\boldsymbol{x}}f(\boldsymbol{x}) + \mu(\boldsymbol{x}, t) ) 
= -\nabla_{\boldsymbol{x}}f(\boldsymbol{x}) - \frac{1}{2} Q(t) \nabla_{\boldsymbol{x}} \ln \rho(\boldsymbol{x}, t).
\label{sec3_eq05}
\end{aligned}
\end{equation}

To verify the correspondence with the Schrödinger equation, we define a quantum state function $\psi : \mathbf{R}^n \times \mathbf{R} \rightarrow \mathbf{C} $ such that
\begin{equation}
\rho(\boldsymbol{x}, t) \triangleq | \psi(\boldsymbol{x}, t) |^2 = \psi(\boldsymbol{x}, t) \cdot \psi^*(\boldsymbol{x}, t) 
\label{sec3_eq06}
\end{equation}
and a correct velocity of a quantum probability current $v : \mathbf{R}^n \times \mathbf{R} \rightarrow \mathbf{C}$ such that 
\begin{equation}
v(\boldsymbol{x}, t) = \frac{\hbar}{i m}(\nabla_{\boldsymbol{x}} \ln \psi(\boldsymbol{x}, t) - \nabla_{\boldsymbol{x}} \ln \psi^*(\boldsymbol{x}, t)) 
\label{sec3_eq07}
\end{equation}
, where $\hbar$ denotes the Dirac constant such that $\hbar = h/2 \pi$ for the Plank constant $h$, $i$ denotes an imaginary unit, $m$ denotes a massive of a particle described by the state $\boldsymbol{x}$, and $\psi^*$ is the conjugate function of $\psi$.

Substituting $\eqref{sec3_eq05}$ and $\eqref{sec3_eq06}$ into $\eqref{sec3_eq03}$, we obtain 
\begin{equation}
\begin{aligned}
&\partial_t \rho(\boldsymbol{x}, t)
= -\nabla \cdot v(\boldsymbol{x}, t) \rho(\boldsymbol{x}, t) \\
&\Rightarrow 
\partial_t \psi^2(\boldsymbol{x}, t)
= -\frac{\hbar}{i m} \nabla \cdot (\nabla_{\boldsymbol{x}} \ln \psi(\boldsymbol{x}, t) - \nabla_{\boldsymbol{x}} \ln \psi^*(\boldsymbol{x}, t)) \psi^2(\boldsymbol{x}, t) \\
&\Rightarrow 
2 \psi(\boldsymbol{x}, t) \partial_t \psi(\boldsymbol{x}, t) 
= -\frac{\hbar}{i m} \nabla \cdot \left( \frac{1}{\psi(\boldsymbol{x}, t) } \nabla_{\boldsymbol{x}} \psi (\boldsymbol{x}, t) 
   - \frac{1}{\psi^*(\boldsymbol{x}, t)} \nabla_{\boldsymbol{x}} \psi^*(\boldsymbol{x}, t) \right) \psi^2(\boldsymbol{x}, t) \\
&\Rightarrow
\partial_t \psi(\boldsymbol{x}, t) 
= -\frac{\hbar}{2 i m} \nabla \cdot \left( \frac{1}{\psi(\boldsymbol{x}, t) } \nabla_{\boldsymbol{x}} \psi(\boldsymbol{x}, t) - \frac{1}{\psi^*(\boldsymbol{x}, t)} \nabla_{\boldsymbol{x}} \psi^*(\boldsymbol{x}, t) \right) \psi(\boldsymbol{x}, t) \\
&\Rightarrow
\partial_t \psi(\boldsymbol{x}, t)
= -\frac{\hbar}{2 i m} \nabla \cdot \left( \nabla_{\boldsymbol{x}} \psi(\boldsymbol{x}, t) - \frac{\nabla_{\boldsymbol{x}} \psi^*(\boldsymbol{x}, t)}{\psi^*(\boldsymbol{x}, t)} \psi(\boldsymbol{x}, t) \right)  \\
&\Rightarrow
i \hbar \partial_t \psi(\boldsymbol{x}, t)
= -\frac{\hbar^2}{2 m} \nabla \cdot \left(\nabla_{\boldsymbol{x}} \psi(\boldsymbol{x}, t) - \frac{\nabla_{\boldsymbol{x}} \psi^*(\boldsymbol{x}, t)}{\psi^*(\boldsymbol{x}, t)} \psi(\boldsymbol{x}, t) \right).
\end{aligned}
\label{sec3_eq08}
\end{equation}

Consequently, if we let a function $V:\mathbf{R}^n \times \mathbf{R} \rightarrow \mathbf{C}$ such that $V(\boldsymbol{x}, t) \triangleq \frac{\hbar^2}{2m} \nabla \cdot \frac{\nabla_{\boldsymbol{x}} \psi^*(\boldsymbol{x}, t)}{\psi^*(\boldsymbol{x}, t)}$, we can obtain the following Schrödinger equation:
\begin{equation}
i \hbar \partial_t \psi(\boldsymbol{x}, t)
= -\frac{\hbar^2}{2 m} \Delta \psi(\boldsymbol{x}, t) + V(\boldsymbol{x}, t) \psi(\boldsymbol{x}, t).
\label{sec3_eq09}    
\end{equation}

\subsection{Quantization parameter in the Quantization-based Optimization from the perspective of the Schrödinger Equation}
In this section, we derive the correspondence of the quantization parameter to the Schrödinger equation $\eqref{sec3_eq09}$.

From the equality of the correct velocity $v(\boldsymbol{x}, t)$ in the equations $\eqref{sec3_eq05}$ and $\eqref{sec3_eq07}$, we can establish the following equation:
\begin{equation}
v(\boldsymbol{x}, t)
= \frac{\hbar}{i m}(\nabla_{\boldsymbol{x}} \ln \psi(\boldsymbol{x}, t) - \nabla_{\boldsymbol{x}} \ln \psi^*(\boldsymbol{x}, t)) 
= -\nabla_{\boldsymbol{x}}f(\boldsymbol{x}) - \frac{1}{2} Q(t) \nabla_{\boldsymbol{x}} \ln \rho(\boldsymbol{x}, t).
\label{sec3_eq10}    
\end{equation}
By the definition of the quantum mechanical probability density in $\eqref{sec3_eq06}$, it leads 
\begin{equation}
\nabla_{\boldsymbol{x}} \ln \rho(\boldsymbol{x}, t) 
=  \nabla_{\boldsymbol{x}} \ln \psi(\boldsymbol{x}, t) \cdot \psi^*(\boldsymbol{x}, t)
=  \nabla_{\boldsymbol{x}} (\ln \psi(\boldsymbol{x}, t) + \ln \psi^*(\boldsymbol{x}, t) ).
\label{sec3_eq11}
\end{equation}
Substituting $\eqref{sec3_eq11}$ into $\eqref{sec3_eq10}$, we get
\begin{equation}
\begin{aligned}
\frac{\hbar}{i m}(\nabla_{\boldsymbol{x}} \ln \psi(\boldsymbol{x}, t) - \nabla_{\boldsymbol{x}} \ln \psi^*(\boldsymbol{x}, t)) 
&= -\nabla_{\boldsymbol{x}}f(\boldsymbol{x}) - \frac{1}{2} Q(t) \nabla_{\boldsymbol{x}} (\ln \psi(\boldsymbol{x}, t) + \ln \psi^*(\boldsymbol{x}, t) ) \\
&= -\nabla_{\boldsymbol{x}}f(\boldsymbol{x}) - \frac{1}{2} Q(t) (\nabla_{\boldsymbol{x}} \ln \psi(\boldsymbol{x}, t) + \nabla_{\boldsymbol{x}} \ln \psi^*(\boldsymbol{x}, t) )
\end{aligned}   
\label{sec3_eq12}    
\end{equation}
If we arrange both terms  by transposition, we obtain
\begin{equation}
\begin{aligned}
-\nabla_{\boldsymbol{x}}f(\boldsymbol{x}) 
&= \frac{\hbar}{i m} \left( \nabla_{\boldsymbol{x}} \ln \psi(\boldsymbol{x}, t) - \nabla_{\boldsymbol{x}} \ln \psi^*(\boldsymbol{x}, t) \right) 
+ \frac{1}{2} Q(t) \left( \nabla_{\boldsymbol{x}} \ln \psi(\boldsymbol{x}, t) + \nabla_{\boldsymbol{x}} \ln \psi^*(\boldsymbol{x}, t) \right) \\
&= Q(t) Re(\nabla_{\boldsymbol{x}} \ln \psi(\boldsymbol{x}, t)) + \frac{2 \hbar}{m} Im(\nabla_{\boldsymbol{x}} \ln \psi(\boldsymbol{x}, t))
\end{aligned}
\label{sec3_eq13}
\end{equation}
Since $\nabla_{\boldsymbol{x}} f(\boldsymbol{x}) \in \mathbf{R}^n$, the equation $\eqref{sec3_eq13}$ is valid.

From the equation $\eqref{sec3_eq12}$, we note that the followings: 
\begin{itemize}
    \item {For a pure deterministic case, i.e., $Q(t)=0$ the gradient of the objective function is proportion to the imaginary part in the gradient of the log scaled quantum state function $\psi$. 
    Thereby, we can describe that the deterministic gradient is a relation to the variation in the frequency of a particle.
    From the viewpoint of numerical analysis, we can regard such a frequency variation based on quantum mechanics as a quantized operation, so that we can describe the deterministic variation of the objective function as the variation of a fundamental power series.  
    }
    \item{For a stochastic case, i.e., $Q(t) > 0$, the gradient of an objective function contains an additional effect of a photon injection.  
    Further,  according to quantum mechanics, we regard $i\hbar \partial_t$ as a total energy $E(\boldsymbol{x}, t)$, and we rewrite $\eqref{sec3_eq09}$ as a following familiar formulation:
    \begin{equation}
    \begin{aligned}
    E(\boldsymbol{x}, t) \psi(\boldsymbol{x}, t) = -\frac{\hbar^2}{2m} \Delta \psi(\boldsymbol{x}, t) + V(\boldsymbol{x}, t) \psi(\boldsymbol{x}, t) \\
    \Rightarrow 
    \Delta \psi(\boldsymbol{x}, t) + \frac{2m}{\hbar^2} \psi(\boldsymbol{x}, t) (E - V)(\boldsymbol{x}, t) = 0.
    \end{aligned}
    \label{sec3_eq14}
    \end{equation}
    In $\eqref{sec3_eq14}$ if we establish a difference energy $U : \mathbf{R}^n \times \mathbf{R} \rightarrow \mathbf{R}$ such that $U = V - E$ and $E < V$ for all $\boldsymbol{x} \in \mathbf{R}^n$ and $t > 0$, we can write $\eqref{sec3_eq14}$ as follows:
    \begin{equation}
    \Delta \psi(\boldsymbol{x}, t) - \frac{2m}{\hbar^2} \psi(\boldsymbol{x}, t) U (\boldsymbol{x}, t) = 0.
    \label{sec3_eq15}
    \end{equation}
    The solution of $\eqref{sec3_eq15}$ reveals that the probability of the state existing beyond the energy hill $V$ is non-zero.
    }
\end{itemize}

In other words, when the current state exists on a local minimum around an energy hill $V$, the stochastic enforcement $Re(\nabla_{\boldsymbol{x}} \ln \psi(\boldsymbol{x}, t))$ controlled by the quantization parameter enables to move the current state on the other state over the energy hill.  
This phenomenon is known as the "tunneling effect."

Accordingly, we note that global optimization techniques such as simulated annealing, quantum annealing, and quantization-based optimization present quantum tunneling, and the hill-climbing based on a noisy vector present equal properties to those of quantum tunneling.

\section{Numerical Experiments}
\begin{table*}[]
\caption{Standard Benchmark Functions}
\begin{tabular}{cc}
\hline
{Function Name}     & {Equation}  \\ 
\hline
{Xin-She Yang N4}   & {$f(x) = 2.0 + (\sum_{i=1}^d \sin^2(x_i) - \exp(-\sum_{i=1}^d x_i^2) \exp(-\sum_{i=1}^d \sin^2 \sqrt{|x_i|})$} \\
{Salomon}           & {$f(x) = 1 - \cos \left(2 \pi \sqrt{\sum_{i=1}^d x_i^2} \right) + 0.1 \sqrt{\sum_{i=1}^d x_i^2}$} \\
{Drop-Wave}         & {$f(x) = 1 - \frac{1 - \cos \left( 12 + \sqrt{x^2 + y^2} \right)}{0.5 (x^2 + y^2) + 2}$} \\
{Shaffel N2}        & {$0.5 + \frac{\sin^2 (x^2 - y^2) - 0.5}{(1 + 0.001(x^2 + y^2)^2}$} \\
\hline
\end{tabular}
\label{result-0}
\end{table*}

\begin{figure*}
    \centering
    \subfigure[Xin-She Yang N4]{
        \centering
        \resizebox{0.23\textwidth}{!}{
        \includegraphics{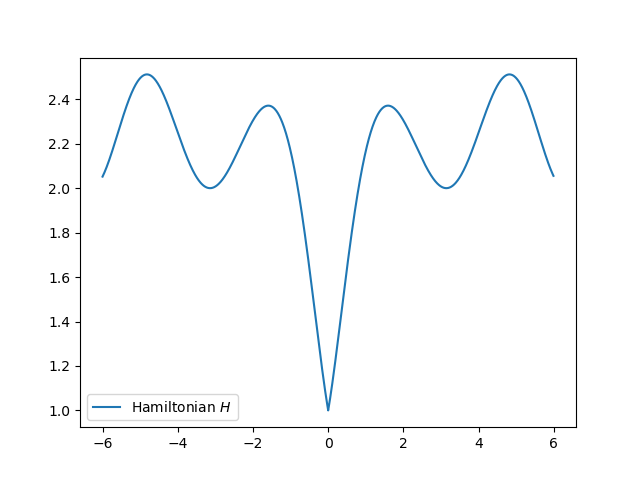}
        }
    }
    \hfill
    \subfigure[Salmon]{
        \centering
        \resizebox{0.23\textwidth}{!}{
        \includegraphics{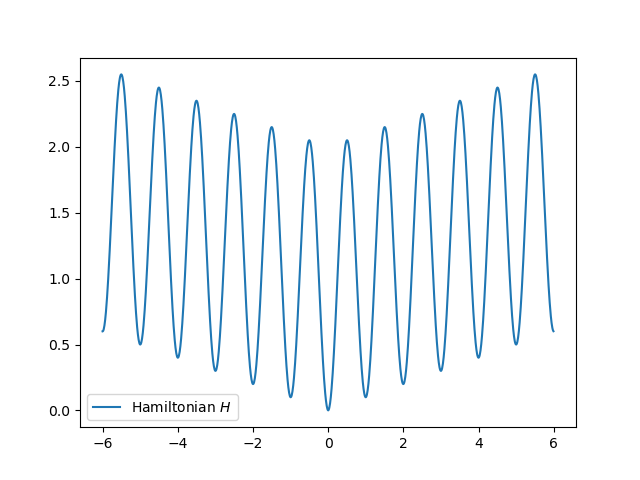}
        }
    }
    \hfill
    \subfigure[Drop-Wave]{
        \centering
        \resizebox{0.23\textwidth}{!}{
        \includegraphics{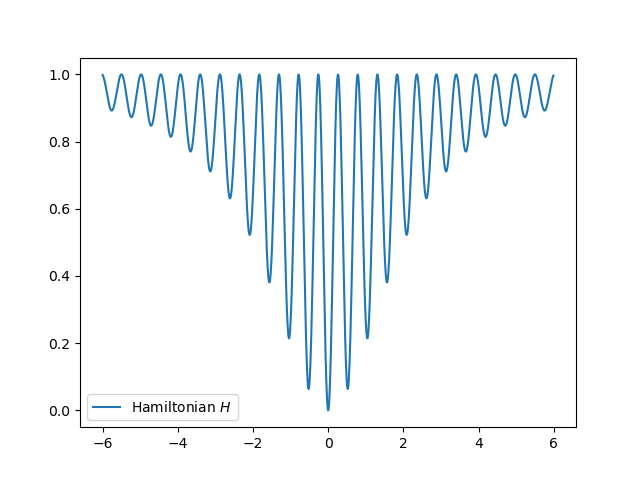}
        }
    }
    \hfill
    \subfigure[Schaffel N2]{
        \centering
        \resizebox{0.23\textwidth}{!}{
        \includegraphics{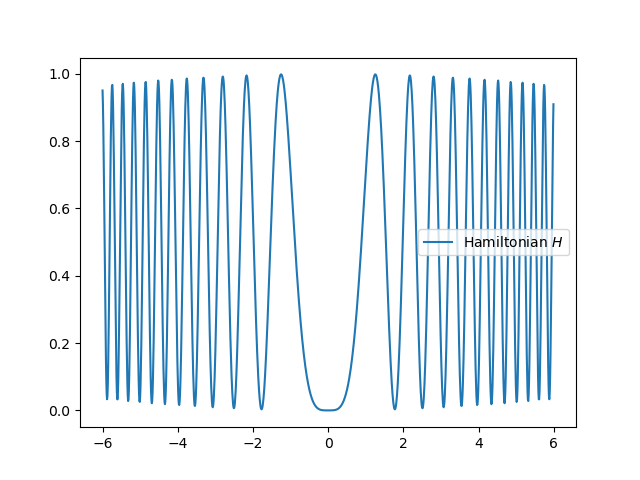}
        }
    }
    \caption{Shape of  Benchmark functions}
    \label{fig:enter-label}
\end{figure*}

To verify the validity of the analysis, we accomplish numerical experiments on optimization problems to multi-modal functions.
The provided benchmark functions are general test functions for optimization algorithms for years. 
In addition, all the benchmark functions contain a lot of local minima along the domain space, so finding the global optimum point in a finite domain is difficult using a conventional deterministic algorithm such as a gradient descent-based optimizer. 
However, as stated in the previous section, if stochastic optimization algorithms, including the quantization-based optimization algorithm, can find the global minimum of the benchmark functions, it reveals that our quantum mechanic-based analysis is valid. 

We employ Simulated Annealing(SA), Quantum Annealing(QA), and the quantization-based optimization algorithm as the stochastic global optimization algorithm for the experiments. 
SA exploited for various combinatorial optimization problems such as the Travelling Salesman Problem (TSP), or Knap-Sack Problem is the representative stochastic global optimization algorithm.  
QA is compatible with combinatorial optimization problems in a similar manner to SA.
In particular, physicists have analyzed the optimization dynamics of QA from the viewpoint of quantum mechanics. 
Quantization-based optimization, which dynamics we analyzed with the perspective of quantum mechanics in this paper, is another type of stochastic optimization for combinatorial optimization. 
Even though SA, QA, and quantization-based optimization are not generally compatible with an optimization problem on a multi-dimensional continuous domain, SA represents sufficient optimization performance on a low-dimensional vector space. 

\begin{table*}[]
\caption{{Simulation results of standard nonlinear optimization functions. SA denotes Simulated Annealing, QA denotes Quantum Annealing, and Quantization represents Quantization-based optimization algorithm}}
\begin{tabular}{clcccc}
\hline
{Function}                         & {Criterion}         & {SA}      & {QA}      & {Quantization} \\ \hline
\multirow{2}{*}{{Xin-She Yang N4}} & {Iteration}         & {6420}    & {17*}     & {3144}         \\
                                   & {Improvement ratio} & {54.57\%} & {35.22\%} & {54.57\%}      \\
\multirow{2}{*}{{Salomon}}         & {Iteration}         & {1312}    & {7092}    & {1727}         \\
                                   & {Improvement ratio} & {99.99\%} & {99.99\%} & {100.0\%}      \\
\multirow{2}{*}{{Drop-Wave}}       & {Iteration}         & {907}     & {3311}    & {254}      \\
                                   & {Improvement ratio} & {100.0\%} & {100.0\%} & {100.0\%}  \\
\multirow{2}{*}{{Shaffel N2}}      & {Iteration}         & {7609}    & {9657}    & {2073}     \\
                                   & {Improvement ratio} & {100.0\%} & {100.0\%} & {100.0\%}  \\ \hline
\end{tabular}
\label{result-1}
\end{table*}

Table \ref{result-1} represents the experimental results. 
As for the Salomon, Drop-wave, and Schaffel N2 benchmark functions, all tested algorithms find the global optimum. 
Those results illustrate the stochastic and quantum mechanics-based optimization algorithms include the same optimization dynamics analyzed with quantum mechanics. 
Further, quantization-based optimization finds the global minimum with fewer iterations than SA and QA. 
This result shows that quantization-based optimization contains an additional property besides the hill-climbing or tunneling effect in optimization.

As for the Xin She Yang N4 benchmark function, the experimental results represent a significantly different aspect. 
SA and quantization-based optimization fall into a local minimum of around 50\% higher value than the global minimum. 
However, the local minimum of the benchmark function is located in a smoother space, whereas the global minimum is located in a very sharp area. 
This result shows that SA and quantization-based algorithms search a minimum point with a positive Hessian with a relatively small matrix norm. 
Practically, the optimization result represents better performance when the algorithm finds such a minimum point in a sparse dataset, whereas the algorithm finding a sharper minimum point occurs as an over-fitting problem.
Finally, in contrast to both algorithms, the QA algorithm fails to find a feasible minimum in the experiment. 
We suppose the reason why QA fails is that the Xin She Yang N4 benchmark function includes a thicker energy barrier to the tunneling effect for searching for a minimum point.

\section{Conclusion}
We present the analysis from the perspective of quantum mechanics for quantization-based optimization in this paper. 
The presented analysis shows that stochastic optimization algorithm, such as quantization-based optimization, includes the tunneling effect to find the global minimum. 
This analysis illustrates that the tunneling effect in quantum mechanical optimization is equal to the hill-climbing property in stochastic algorithms. 
Finally, in future work, we will research the hidden dynamics of why quantization-based optimization represents finding the global minimum with fewer iterations. 

\section*{Acknowledgment}
This work was supported by the Institute for Information Communications Technology PlanningEvaluation(IITP) grant funded by the Korean government(MSIT) (No.2021-0-00766 Development of Integrated Development Framework that supports Automatic Neural Network Generation and Deployment optimized for Runtime Environment).

\newpage
\bibliography{arxiv_main}

\end{document}